%% file: 071001arXivSubmit.tex
\documentclass[]{iopart}
\input amssym.def
\input amssym
\usepackage{graphicx}

\begin{document}

\input{mycommandsSQM.tex}

\title{pQCD vs.\ AdS/CFT Tested by Heavy Quark Energy Loss}

\author{W.\ A.\ Horowitz}

\address{Department of Physics, 
Columbia University,
538 West 120$^\mathrm{th}$ Street,
New York, NY 10027, USA}

\address{Frankfurt Institute for Advanced Study (FIAS),
60438 Frankfurt am Main, Germany}

\ead{horowitz@phys.columbia.edu}

\begin{abstract}
We predict the charm and bottom quark nuclear modification factors 
using weakly coupled pQCD and strongly coupled \ads drag methods.  The $\log(\eqnpt/M_Q)/\eqnpt$ dependence of pQCD loss and the momentum independence of drag loss lead to different momentum dependencies for the \raa predictions.  This difference is enhanced by examining a new experimental observable, the double ratio of charm to bottom nuclear modification factors, 
$R^{cb}=R_{AA}^c/R_{AA}^b$.  At LHC the weakly coupled theory predicts $R^{cb}\rightarrow1$ whereas the strongly coupled theory predicts $R^{cb} \sim .2$ independent of \ptcomma.  At RHIC the differences are less dramatic, as the production spectra are harder, but the drag formula is applicable to higher momenta, due to the lower temperature.
\end{abstract}
\vspace{-.2in}
\vspace{-.27in}
\submitto{\jpg}
\vspace{-.13in}
\section{Introduction}
The excellent agreement between the pion nuclear modification factor, $R_{AA}^\pi(\eqnpt)$, and the perturbative quantum chromodynamics (pQCD) prediction, combined with the null control experiments of $d+A$, direct photons, and the consistency check of $R_{AA}^\pi(\eqnpt) \sim R_{AA}^\eta(\eqnpt)$ \cite{Akiba:2005bs}, makes a compelling argument for the creation of a weakly coupled quark-gluon plasma (QGP) at RHIC.  However this picture of a weakly coupled plasma is undermined by the failure of pQCD to simultaneously account for several other experimental observables: the nearly ideal hydrodynamic flow of \lowpt particles (requiring an entropy to viscosity ratio $\eta/s \sim .1$); the highly suppressed (similar in magnitude to the pions) nonphotonic electrons, the decay fragments of heavy charm and bottom quarks; and the nonperturbatively large \intermediatept azimuthal anisotropy, \vtwo \cite{Teaney:2003kp}.  Much progress has been made in reducing the discrepancies between pQCD and data; one cannot currently claim the falsification of pQCD techniques applied to RHIC \cite{El:2007eh}.  Nonetheless, one might naturally consider the possibility that strongly coupled dynamics might provide a better theoretical description of data.
%

Exploiting the AdS/CFT correspondence, which allows many strongly coupled supersymmetric field theoretic problems to be solved analytically in a classical supergravity (SUGRA) dual, a number of calculations have been made that are, in fact, in qualitative agreement with known results.  The infinitely coupled entropy density is three-fourths the value of the Stefan-Boltzmann limit, similar to that seen on the lattice.  There appears to be a universal lower bound of entropy to viscosity, $\eta/s \geq 1/4\pi \sim.1$.  Mach wave-like structures appear for supersonic motion.  And heavy quarks moving through strongly coupled plasma appear to lose a significant fraction of their momentum \cite{Kovtun:2004de}.  Despite these successes, super Yang-Mills (SYM) is quite different from QCD and, partly for this reason, strong coupling calculations have so far failed to make precision predictions.

What we seek, then, is a robust, distinguishing measurement that could possibly falsify one or both approaches.  It turns out that measuring the heavy quark energy loss for charm and bottom quarks separately provides just such a probe.  

\section{AdS/CFT Energy Loss Compared to pQCD}
The AdS/CFT conjecture has been applied to three types of energy loss models.  The first uses the Wilson line formulation for the radiative transport coefficient \qhatcomma; the conjecture allows a numerical estimate of its value.  The second finds the momentum diffusion coefficient, $D$, used in a collisional heavy quark energy loss computation.  The third models the entire plasma plus heavy quark system in the AdS/CFT dual and derives the drag coefficient \cite{Liu:2006ug}.  
There is considerable ongoing debate and research on these approaches (see, e.g.\ \cite{Liu:2006ug,Liu:2006he}); we will focus on the latter because it does not use the AdS/CFT correspondence to find the value of an input parameter needed in a pQCD-inspired model.  Rather its more ambitious goal of modeling the entire energy loss process in the SUGRA dual leads to novel mass and momentum dependencies.

The drag on a heavy quark of mass $M_Q$ moving with constant velocity in a constant temperature SYM plasma in the limit of $\lambda = \sqrt{g_{SYM}^2N_c} \gg 1$, $N_c \gg 1$, $M_Q \gg T^{SYM}$ is \cite{Liu:2006ug}
\be
\label{mu}
\frac{d\eqnpt}{dt} = -\mu_Q\eqnpt = \frac{\pi\sqrt{\lambda}(T^{SYM})^2}{2M_Q}\eqnpt.
\ee
Issues related to the relaxation of the strong assumptions made in deriving \eq{mu} and the momenta above which one begins to expect significant corrections are discussed later in the text.

One typically does not consider drag, but rather fractional energy loss $\epsilon$, where $p_f = (1-\epsilon) p_i$.  For approximately constant, luminal velocity the average fractional loss for a quark propagating through a medium of length $L$ is
$\bar{\epsilon}_{AdS} = 1 - \exp(-\mu_Q L)$, \emph{independent} of \ptcomma.  On the other hand the asymptotic form of the average pQCD radiative fractional energy loss is
$\bar{\epsilon}_{pQCD} = \kappa \eqnalphas \eqnqhat L^2 \log(\eqnpt/M_Q)/\eqnpt$, where $\kappa$ is a dimensionless proportionality constant.  In this case $\bar{\epsilon}_{pQCD}\rightarrow0$ as \pt increases.  

The exceptional momentum reach of LHC and its particle identification (PID) capabilities can be used to distinguish between these two.  The production spectrum for heavy quarks is well approximated by a power law, $dN_Q/d\eqnpt \sim 1/\eqnpt^{n_Q(\eqnpt)}$, with $dn_Q/d\eqnpt > 0$.  We then have
$R_{AA}^{Q}(\eqnpt) = \langle \left( 1-\epsilon(\vec{x},\phi) \right)^{n_Q(\eqnpt)-1} \rangle_{geom}$, where, in the full numerical results shown in the figures the average over realistic geometry includes jet production proportional to the binary distribution and propagation through a medium whose transverse profile is proportional to the participant distribution.  Bjorken expansion is included, and a  diffuse Woods-Saxon base nuclear profile is used.  $n_Q(\eqnpt)$ was found from a best fit of FONLL spectra \cite{Horowitz:2007su}.

For LHC the increase in the power law index is not fast enough to compensate for the decrease in pQCD fractional energy loss; thus $dR_{AA}^{pQCD}/d\eqnpt > 0$.  Contrariwise, the increase in the index combined with the momentum independence of the \ads drag implies $dR_{AA}^{AdS}/d\eqnpt < 0$.  

To compare the AdS/CFT drag predictions to pQCD predictions and future experimental data one must connect SYM to QCD.  This requires a nontrivial mapping of the running coupling in QCD to the constant coupling in SYM and of the temperatures.  The most na\"ive we denote as ``obvious''; this takes $\eqnalphasym = \eqnalphas = const.$ and $T^{SYM} = T^{QCD}$.  An ``alternate'' might equate the quark-antiquark force from AdS/CFT to that measured on the lattice, yielding $\lambda_{SYM} \simeq 5.5$, and the energy densities, yielding, due to the approximately factor of three greater number of degrees of freedom in SYM, $T^{SYM} = T^{QCD}/3^{1/4}$ \cite{Liu:2006ug}.  

As an illustration of one of the difficulties in making a precision strong coupling prediction consider the momentum diffusion coefficient mentioned earlier, $D=2/\sqrt{\lambda}\pi T$ \cite{Liu:2006ug}.  Using the ``obvious'' prescription and $\eqnalphas = .3$, a common value taken in pQCD jet physics \cite{El:2007eh}, $D \simeq 1.2/2\pi T$.  However $D \simeq 3/2\pi T$ was claimed to better reproduce experimental data \cite{Liu:2006ug}.  This would require a coupling of $\eqnalphas\simeq.05$, far from the strong coupling limit.  Nevertheless we will use these two $\eqnalphas$'s as representative values when employing the ``obvious'' prescription.  We note that a smaller coupling leads to a larger diffusion coefficient because $D$ is in the coordinate representation instead of the usual momentum one.  

\eq{mu} is implemented numerically as in \cite{Horowitz:2007su}.  Energy loss is assumed to start at thermalization, $\tau_0 \sim .6-1$ fm/c, and stops when the confinement temperature, $T_c \sim 160$ MeV, is reached.  The exponentiated $T^2$ dependence in $\mu_Q$ leads to an extremely sensitive probe of the opacity of the medium, as well as $\tau_0$ and $T_c$.  %
We will employ the WHDG model of convolved elastic and radiative energy loss for pQCD predictions \cite{El:2007eh}.  
LHC predictions using the ASW model of purely radiative energy loss have also been made with extrapolations of extreme $40 \lesssim \eqnqhat \lesssim 100$ \cite{Armesto:2003jh}; instead of using the exact ASW model we approximate these predictions with radiative only WHDG.  
In all calculations initial state effects were neglected; just as at RHIC for definitive experimental statements to be made a $p+Pb$ control run will be essential.

\begin{figure}[htb!]
\vspace{-.05in}
\begin{center}
\leavevmode
\includegraphics[viewport = 2 -1 933 466, clip = true, width=.75\columnwidth]{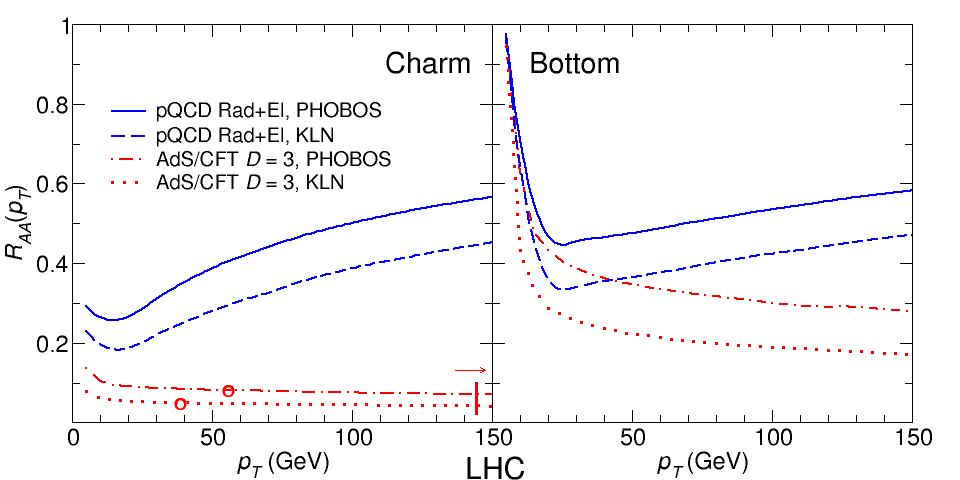}
\vspace{-.15in}
\caption{
\label{fig1}
\captionsize{(Color Online) \raacpt and \raabpt for central $Pb+Pb$ reactions at LHC with energy loss given by AdS/CFT drag and by pQCD using WHDG \cite{El:2007eh}.  The full numerics produce the expected behavior of $dR_{AA}^{AdS}/d\eqnpt < 0$ and $dR_{AA}^{pQCD}/d\eqnpt > 0$.  Representative possible initial gluon densities are given by $\eqndngslashdy = 1750$ and $\eqndngslashdy = 2900$ from the PHOBOS extrapolation and the KLN model of the CGC, respectively \cite{Back:2001ae}.
The \ads drag parameters are from the ``obvious'' prescription with $\alpha_{SYM}=.05$, giving $D = 3/2\pi T$ (abbreviated to $D=3$ in the figure).  The ``O'' and ``$|$'' symbols come from \eq{speedlimit} and are explained in the text.}
}
\end{center}
\vspace{-.2in}
\end{figure}

In \fig{fig1} we see that our expectations for the momentum dependence of the charm and bottom \raa for the two theories are borne out in the full numerical calculations.  While not shown, one is not surprised that the purely radiative, highly suppressed, large \qhat extrapolations have momentum dependencies difficult to distinguish from some of the \ads drag predictions.  It turns out that we can enhance the signal by taking the \emph{double} ratio $R^{cb}(\eqnpt) \equiv \eqnraacpt/\eqnraabpt$.  From the asymptotic fractional energy loss formula above pQCD gives
\be
R^{cb}_{pQCD}(\eqnpt) \simeq 1 - \kappa \eqnalphas n(\eqnpt) L^2 \log(M_b/M_c) \frac{\eqnqhat}{\eqnpt},
\ee
where $n_b(\eqnpt) \simeq n_c(\eqnpt) \simeq n(\eqnpt)$, and the argument of the $\log$ is the ratio of bottom to charm quark mass.  Hence pQCD predicts that this ratio goes to unity asymptotically, and that this approach is slower for greater suppression.  

Taking the medium to be have a uniform, time-independent density of size $L$, one may estimate the \ads drag nuclear modification factor: $R_{AA}^{Q} \sim 1/n_Q\mu_QL$.  Then
\be
R^{cb}_{AdS}(\eqnpt) \simeq \frac{n_b\mu_b}{n_c\mu_c} \simeq \frac{M_c}{M_b},
\ee
and \ads drag predicts $R^{cb}\approx.27$, independent of momentum.  

\begin{figure}[htb!]
\vspace{-.15in}
\begin{center}
\leavevmode
\includegraphics[viewport = 0 0 1283 729, clip = true, width=.75\columnwidth]{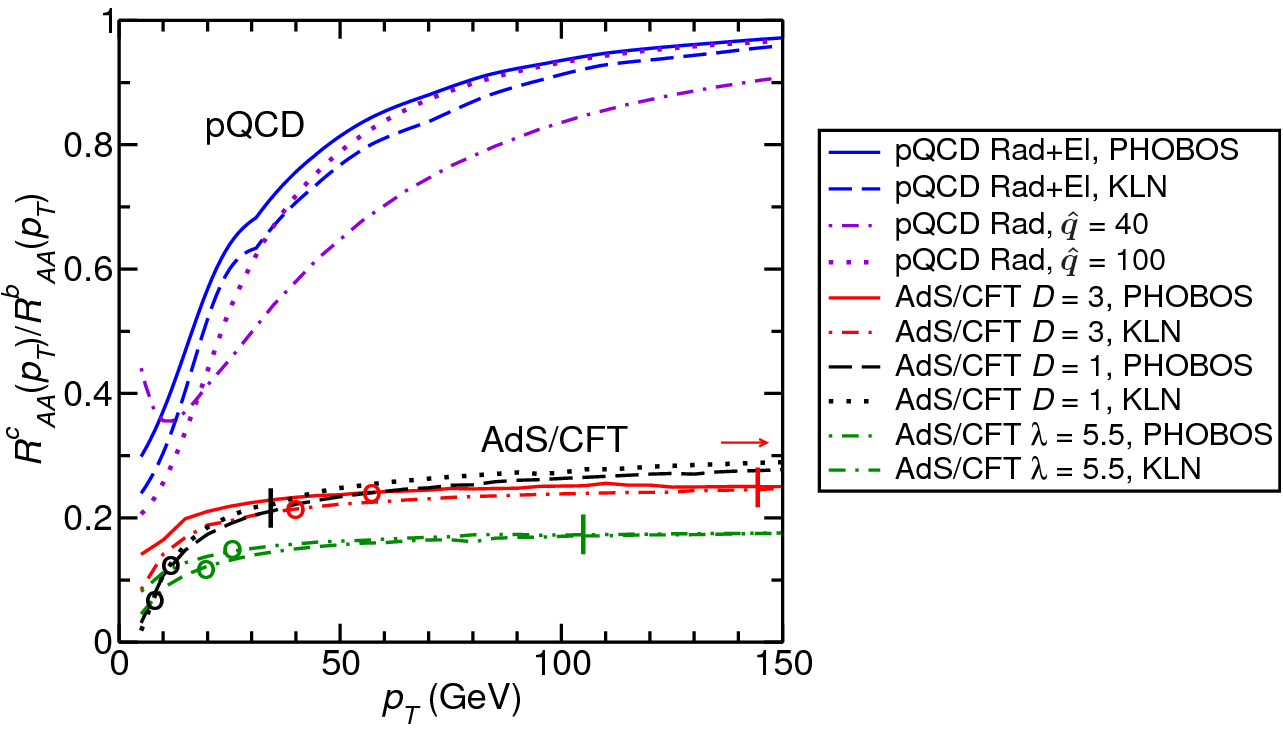}
\vspace{-.15in}
\caption{
\label{ratio}
\captionsize{The double ratio of \raacpt to \raabpt predictions for LHC using \eq{mu} for \ads and WHDG \cite{El:2007eh} for pQCD with a wide range of input parameters.  By taking the ratio the large normalization differences cancel, and the results collect into two distinct groups.  }
}
\end{center}
\vspace{-.2in}
\end{figure}

One sees in \fig{ratio} that the above approximations are well reproduced by the full numerical results.  Most normalization differences for the wide range of input parameters used cancel, and the curves bunch into a pQCD group and an \ads drag group.  Additionally the more highly quenched pQCD curves approach 1 more slowly, except for the highly oversuppressed $\eqnqhat = 100$.  Supposing that LHC data are similar to the pQCD predictions one might be able to distinguish between convolved elastic and inelastic loss, for which $R^{cb}$ monotonically increases, and purely radiative energy loss, for which $R^{cb}$ dips to a minimum near $\eqnpt \sim 10$ GeV; one must be cautious, though, as many of the assumptions in the elastic energy loss derivations break down for $\eqnpt\not\gg M_Q$.

It turns out, though, that there is reason to believe that \eq{mu} cannot be applied to arbitrarily large \ptcomma.  For strictly finite heavy quark mass, demanding a time-like string endpoint in constant velocity motion gives a ``speed limit'' of \cite{Gubser:2006nz}
\be
\label{speedlimit}
\gamma_c=\left(1+\frac{2M}{\sqrt{\lambda}T^{SYM}}\right)^2\approx\frac{4M^2}{\lambda(T^{SYM})^{2}}.
\ee
One may arrive at this same value through a different line of reasoning.  Assume that the quark's constant velocity is maintained by an electric field.  The largest electric field sustainable from the Born-Infeld action limits the magnitude of the momentum loss; the critical speed after which the field cannot be strong enough to keep the quark velocity constant is given by \eq{speedlimit} \cite{Gubser:2006nz}.  

\eq{speedlimit} was found assuming a constant plasma temperature; this is not the case in experiment.  To give some idea of the momenta above which one might expect significant corrections to \eq{mu} we plot the smallest possible $\gamma_c$ (corresponding to the largest temperature, $T(\vec{x}=0,t=\tau_0$) and the largest $\gamma_c$ (from the smallest temperature, $T_c$).  These are indicated by an ``O'' and a ``$|$'' in the figures, respectively.

\begin{figure}[htb!]
\begin{center}
\leavevmode
$\begin{array}{c@{\hspace{.05in}}c}
\includegraphics[viewport = 20 25 423 416, clip = true, width=.32\columnwidth]{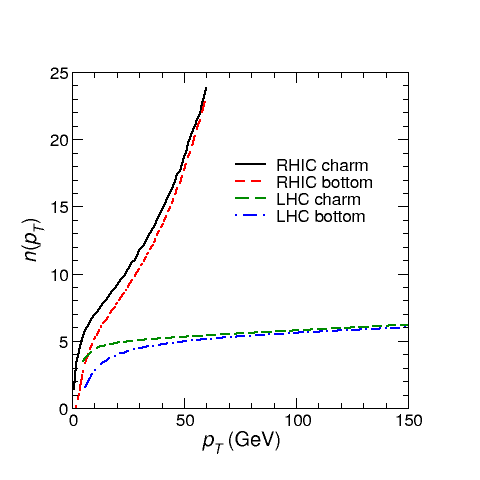} & 
\includegraphics[viewport = 2 0 950 466, clip = true, width=.63\columnwidth]{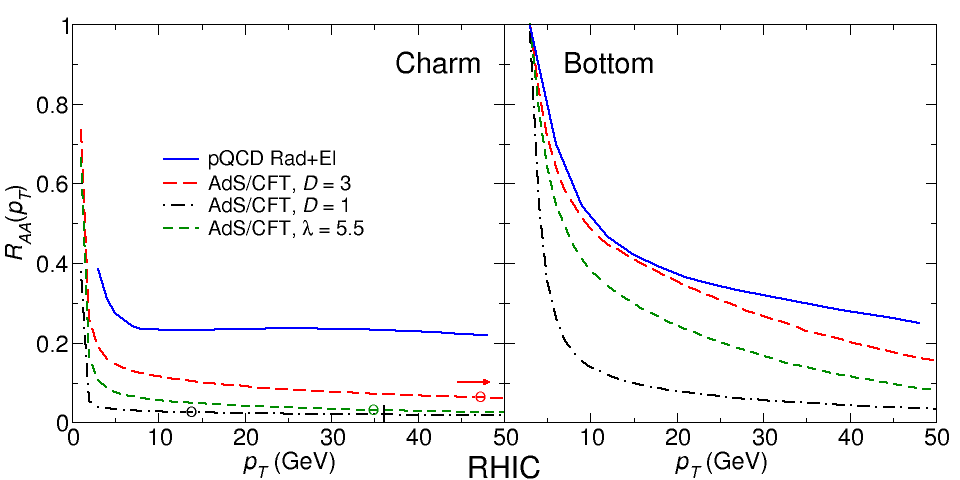} \\ [-0.in]
{\mbox {\scriptsize {\bf (a)}}} & {\mbox {\scriptsize {\bf (b)}}}
\end{array}$
\vspace{-.15in}
\caption{
\label{rhic}\captionsize{(a) The power law production index $n_Q(\eqnpt)$ for RHIC and LHC.  While it is quite flat over a large momentum range at LHC, at RHIC $n_Q(\eqnpt)$ hardens appreciably as momentum increases. (b) \raacpt and \raabpt for RHIC using \ads drag and pQCD WHDG.  The large increase in $n_Q(\eqnpt)$ overcomes the decreasing fractional momentum loss for pQCD; both \ads drag and pQCD results decrease as a function of momentum at RHIC.}
}
\end{center}
\vspace{-.2in}
\end{figure}

\begin{figure}[htb!]
\begin{center}
\leavevmode
\includegraphics[width=.75\columnwidth]{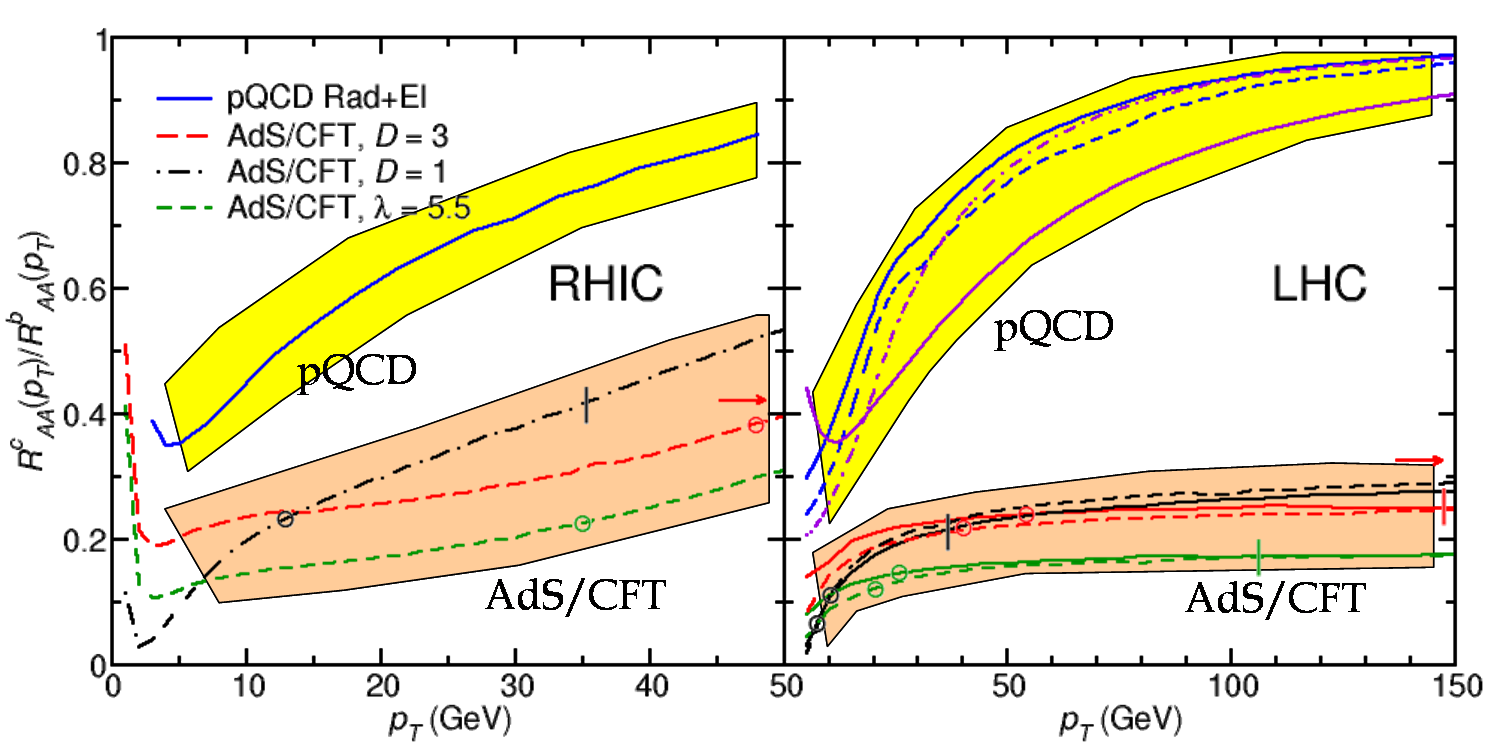}
\vspace{-.15in}
\caption{
\label{rhicratio}
\captionsize{The double ratio of \raacpt to \raabpt predictions for RHIC using \eq{mu} for \ads and WHDG \cite{El:2007eh} for pQCD with a range of input parameters.  While the hardening of the production spectrum reduces the dramatic bunching at RHIC as compared to LHC, the lower temperature at RHIC means the \ads drag formalism is applicable to higher momenta.  Note that $R^{cb}$ is plotted to only 50 GeV for RHIC.}
}
\end{center}
\vspace{-.2in}
\end{figure}

Future RHIC detector upgrades will allow for individual charm and bottom quark detection.  Predictions for \raacpt and \raabpt at RHIC from \eq{mu} and pQCD are shown in \fig{rhic} (b).  As seen in \fig{rhic} (a) the power law production index grows quickly at RHIC; we no longer can expect that the decrease in pQCD energy loss as a function of momentum will result in $dR_{AA}^{pQCD}/d\eqnpt > 0$.  In fact \fig{rhic} (b) shows that the full numerical results for $R_{AA}^Q$ from pQCD and \ads drag both decrease with \ptcomma.  Nonetheless one may still examine the double ratio \rcbcomma, \fig{rhicratio}.  While the larger index makes the grouping less dramatic at RHIC one may still differentiate between pQCD and \ads drag.  Due to its smaller multiplicities, the temperature of the medium at RHIC is smaller than will be seen at LHC; hence the \ads drag ``speed limit'', \eq{speedlimit}, is higher at RHIC than LHC.  

\vspace{-.2in}
\section{Conclusions}
We compared the $R_{AA}^Q(\eqnpt)$ predictions from weakly coupled pQCD and strongly coupled AdS/CFT.  The momentum dependencies of the energy loss formulae resulted in $dR_{AA}^{pQCD}/d\eqnpt > 0$ and $dR_{AA}^{AdS}/d\eqnpt < 0$ at LHC.  The difference in the mass and momentum scales for the two make the new observable $R^{cb}(\eqnpt) = \eqnraacpt / \eqnraabpt$ an excellent probe for falsifying a coupling limit.  
These calculations were also performed for RHIC.  The harder power law index meant that both pQCD and \ads drag predicted decreasing $dR_{AA}^Q/d\eqnpt < 0$.  The double ratio $R^{cb}$, though, still had differentiating power.  Two issues are of concern: momentum range of applicability and universality.  pQCD calculations may only be valid at very high \ptcomma; \ads drag calculations may only be valid up to some lower momentum limit.  It is not impossible that an overlap region for which both ought to be valid does not exist.  The possibility of falsification rests with unique, measurable differences in theoretical predictions.  A theory which does not give this may not be falsified by the data, but is also of little use.

\end{document}

%% file: mycommandsSQM.tex
\newcommand{\be}{\begin{equation}}
\newcommand{\ee}{\end{equation}}
\newcommand{\bfig}{\begin{figure}}
\newcommand{\efig}{\end{figure}}
\newcommand{\bea}{\begin{eqnarray}}
\newcommand{\eea}{\end{eqnarray}}
\newcommand{\infinitessimal}{\mathrm{d}}
\newcommand{\infinitesmal}{\mathrm{d}}
\newcommand{\infinitesimal}{\mathrm{d}}
\newcommand{\intd}{\mathrm{d}}


\newcommand{\raa}{$R_{AA}$ }
\newcommand{\raacomma}{$R_{AA}$}
\newcommand{\raaphi}{$R_{AA}(\phi)$ }
\newcommand{\raaphicomma}{$R_{AA}(\phi)$}
\newcommand{\raaphipt}{$R_{AA}(\phi;\,\eqnpt)$ }
\newcommand{\raaphiptcomma}{$R_{AA}(\phi;\,\eqnpt)$} 
\newcommand{\raapt}{$R_{AA}(\eqnpt)$ } 
\newcommand{\raaptcomma}{$R_{AA}(\eqnpt)$} 
\newcommand{\eqnraapt}{R_{AA}(\eqnpt)} 
\newcommand{\raaq}{$R_{AA}^q$ }
\newcommand{\raaqcomma}{$R_{AA}^q$}
\newcommand{\eqnraaq}{R_{AA}^q}
\newcommand{\raaqphi}{$R_{AA}^q(\phi)$ }
\newcommand{\raaqphicomma}{$R_{AA}^q(\phi)$} 
\newcommand{\eqnraaqphi}{R_{AA}^q(\phi)}
\newcommand{\raaqphipt}{$R_{AA}^q(\phi;\,\eqnpt)$ }
\newcommand{\raaqphiptcomma}{$R_{AA}^q(\phi;\,\eqnpt)$} 
\newcommand{\eqnraaqphipt}{R_{AA}^q(\phi;\,\eqnpt)} 
\newcommand{\raaqpt}{$R_{AA}^q(\eqnpt)$ } 
\newcommand{\raaqptcomma}{$R_{AA}^q(\eqnpt)$} 
\newcommand{\eqnraaqpt}{R_{AA}^q(\eqnpt)}
\newcommand{\mq}{$m_q$ }
\newcommand{\mqcomma}{$m_q$}
\newcommand{\eqnmq}{m_q}
\newcommand{\muq}{$\mu_q$ }
\newcommand{\muqcomma}{$\mu_q$}
\newcommand{\eqnmuq}{\mu_q}
 
\newcommand{\raaQ}{$R_{AA}^Q$ }
\newcommand{\raaQcomma}{$R_{AA}^Q$}
\newcommand{\eqnraaQ}{R_{AA}^Q}
\newcommand{\raaQphi}{$R_{AA}^Q(\phi)$ }
\newcommand{\raaQphicomma}{$R_{AA}^Q(\phi)$} 
\newcommand{\eqnraaQphi}{R_{AA}^Q(\phi)}
\newcommand{\raaQphipt}{$R_{AA}^Q(\phi;\,\eqnpt)$ }
\newcommand{\raaQphiptcomma}{$R_{AA}^Q(\phi;\,\eqnpt)$} 
\newcommand{\eqnraaQphipt}{R_{AA}^Q(\phi;\,\eqnpt)} 
\newcommand{\raaQpt}{$R_{AA}^Q(\eqnpt)$ } 
\newcommand{\raaQptcomma}{$R_{AA}^Q(\eqnpt)$} 
\newcommand{\eqnraaQpt}{R_{AA}^Q(\eqnpt)}
\newcommand{\mQ}{$m_Q$ }
\newcommand{\mQcomma}{$m_Q$}
\newcommand{\eqnmQ}{m_Q}
\newcommand{\muQ}{$\mu_Q$ }
\newcommand{\muQcomma}{$\mu_Q$}
\newcommand{\eqnmuQ}{\mu_Q}

\newcommand{\raac}{$R_{AA}^c$ }
\newcommand{\raaccomma}{$R_{AA}^c$}
\newcommand{\eqnraac}{R_{AA}^c}
\newcommand{\raacphi}{$R_{AA}^c(\phi)$ }
\newcommand{\raacphicomma}{$R_{AA}^c(\phi)$} 
\newcommand{\eqnraacphi}{R_{AA}^c(\phi)}
\newcommand{\raacphipt}{$R_{AA}^c(\phi;\,\eqnpt)$ }
\newcommand{\raacphiptcomma}{$R_{AA}^c(\phi;\,\eqnpt)$} 
\newcommand{\eqnraacphipt}{R_{AA}^c(\phi;\,\eqnpt)} 
\newcommand{\raacpt}{$R_{AA}^c(\eqnpt)$ } 
\newcommand{\raacptcomma}{$R_{AA}^c(\eqnpt)$} 
\newcommand{\eqnraacpt}{R_{AA}^c(\eqnpt)} 

\newcommand{\raab}{$R_{AA}^b$ }
\newcommand{\raabcomma}{$R_{AA}^b$}
\newcommand{\eqnraab}{R_{AA}^b}
\newcommand{\raabphi}{$R_{AA}^b(\phi)$ }
\newcommand{\raabphicomma}{$R_{AA}^b(\phi)$} 
\newcommand{\eqnraabphi}{R_{AA}^b(\phi)}
\newcommand{\raabphipt}{$R_{AA}^b(\phi;\,\eqnpt)$ }
\newcommand{\raabphiptcomma}{$R_{AA}^b(\phi;\,\eqnpt)$} 
\newcommand{\eqnraabphipt}{R_{AA}^b(\phi;\,\eqnpt)} 
\newcommand{\raabpt}{$R_{AA}^b(\eqnpt)$ } 
\newcommand{\raabptcomma}{$R_{AA}^b(\eqnpt)$} 
\newcommand{\eqnraabpt}{R_{AA}^b(\eqnpt)}

\newcommand{\cbratio}{$\eqnraacpt/\eqnraabpt$ }
\newcommand{\cbratiocomma}{$\eqnraacpt/\eqnraabpt$}
\newcommand{\eqncbratio}{\eqnraacpt/\eqnraabpt}
\newcommand{\rcb}{$R^{cb} $}
\newcommand{\rcbcomma}{$R^{cb}$}
\newcommand{\eqnrcb}{R^{cb}}

\newcommand{\raag}{$R_{AA}^g$ }
\newcommand{\raagcomma}{$R_{AA}^g$}
\newcommand{\eqnraag}{R_{AA}^g}
\newcommand{\raagphi}{$R_{AA}^g(\phi)$ }
\newcommand{\raagphicomma}{$R_{AA}^g(\phi)$} 
\newcommand{\eqnraagphi}{R_{AA}^g(\phi)}
\newcommand{\raagphipt}{$R_{AA}^g(\phi;\,\eqnpt)$ }
\newcommand{\raagphiptcomma}{$R_{AA}^g(\phi;\,\eqnpt)$} 

\newcommand{\eqnraagphipt}{R_{AA}^g(\phi;\,\eqnpt)} 
\newcommand{\raagpt}{$R_{AA}^g(\eqnpt)$ } 
\newcommand{\raagptcomma}{$R_{AA}^g(\eqnpt)$} 
\newcommand{\eqnraagpt}{R_{AA}^g(\eqnpt)} 

\newcommand{\RAA}{\raa}
\newcommand{\RAAcomma}{\raacomma}
\newcommand{\RAAphi}{\raaphi}
\newcommand{\RAAphicomma}{\raaphicomma}
\newcommand{\RAAphipt}{\raaphipt}
\newcommand{\RAAphiptcomma}{\raaphiptcomma}
\newcommand{\raapi}{$R_{AA}^\pi$ }
\newcommand{\raae}{$R_{AA}^{e^-}$ }
\newcommand{\raapicomma}{$R_{AA}^\pi$}
\newcommand{\raaecomma}{$R_{AA}^{e^-}$}
\newcommand{\eqnraapi}{R_{AA}^\pi}
\newcommand{\eqnraae}{R_{AA}^{e^-}}

\newcommand{\vtwo}{$v_2$ }
\newcommand{\vtwocomma}{$v_2$}
\newcommand{\vtwopt}{$v_2(\eqnpt)$ }
\newcommand{\vtwoptcomma}{$v_2(\eqnpt)$}
\newcommand{\eqnraa}{R_{AA}}
\newcommand{\eqnraaphi}{R_{AA}(\phi)}
\newcommand{\eqnraaphipt}{R_{AA}(\phi;\,\eqnpt)} 
\newcommand{\eqnRAA}{\eqnraa}
\newcommand{\eqnRAAphi}{\eqnraaphi}
\newcommand{\eqnRAAphipt}{\eqnraaphipt}
\newcommand{\eqnvtwo}{v_2}
\newcommand{\vtwovsraa}{\vtwo vs.~\raa}
\newcommand{\vtwovsraacomma}{\vtwo vs.~\raacomma}
\newcommand{\eqnvtwopt}{v_2(\eqnpt)}

\newcommand{\pp}{$p+p$ }
\newcommand{\ppcomma}{$p+p$}
\newcommand{\dau}{$d+Au$ }
\newcommand{\daucomma}{$d+Au$}
\newcommand{\auau}{$Au+Au$ }
\newcommand{\auaucomma}{$Au+Au$}
\newcommand{\aplusa}{$A+A$ }
\newcommand{\aplusacomma}{$A+A$}
\newcommand{\cucu}{$Cu+Cu$ }
\newcommand{\cucucomma}{$Cu+Cu$}

\newcommand{\rhopart}{$\rho_{\textrm{\footnotesize{part}}}$ }
\newcommand{\rhopartcomma}{$\rho_{\textrm{\footnotesize{part}}}$}
\newcommand{\eqnrhopart}{\rho_{\textrm{\footnotesize{part}}}}
\newcommand{\npart}{$N_{\textrm{\footnotesize{part}}}$ }
\newcommand{\npartcomma}{$N__{\textrm{\footnotesize{part}}}$}
\newcommand{\eqnnpart}{N_{\textrm{\footnotesize{part}}}}
\newcommand{\taa}{$T_{AA}$ }
\newcommand{\taacomma}{$T_{AA}$}
\newcommand{\eqntaa}{T_{AA}}
\newcommand{\rhocoll}{$\rho_{\textrm{\footnotesize{coll}}}$ }
\newcommand{\rhocollcomma}{$\rho_{\textrm{\footnotesize{coll}}}$}
\newcommand{\eqnrhocoll}{\rho_{\textrm{\footnotesize{coll}}}}
\newcommand{\ncoll}{$N_{\textrm{\footnotesize{coll}}}$ }
\newcommand{\ncollcomma}{$N_{\textrm{\footnotesize{coll}}}$}
\newcommand{\eqnncoll}{N_{\textrm{\footnotesize{coll}}}}
\newcommand{\dndy}{$\frac{dN_g}{dy}$ }
\newcommand{\dndycomma}{$\frac{dN_g}{dy}$}
\newcommand{\eqndndy}{\frac{dN_g}{dy}}
\newcommand{\eqndndyabs}{\frac{dN_g^{abs}}{dy}}
\newcommand{\eqndndyrad}{\frac{dN_g^{rad}}{dy}}
\newcommand{\dnslashdy}{$dN_g/dy$ }
\newcommand{\dnslashdycomma}{$dN_g/dy$}
\newcommand{\eqndnslashdy}{dN_g/dy}
\newcommand{\dngdy}{$\frac{dN_g}{dy}$ }
\newcommand{\dngdycomma}{$\frac{dN_g}{dy}$}
\newcommand{\eqndngdy}{\frac{dN_g}{dy}}
\newcommand{\eqndngdyabs}{\frac{dN_g^{abs}}{dy}}
\newcommand{\eqndngdyrad}{\frac{dN_g^{rad}}{dy}}
\newcommand{\dngslashdy}{$dN_g/dy$ }
\newcommand{\dngslashdycomma}{$dN_g/dy$}
\newcommand{\eqndngslashdy}{dN_g/dy}
\newcommand{\as}{\alpha_s}
\newcommand{\alphas}{$\as$ }
\newcommand{\alphascomma}{$\as$}
\newcommand{\eqnalphas}{\as}

\renewcommand{\pt}{$p_T$ }
\newcommand{\pT}{\pt}
\newcommand{\ptcomma}{$p_T$}
\newcommand{\pTcomma}{\ptcomma}
\newcommand{\eqnpt}{p_T}
\newcommand{\ptf}{$p_T^f$ }
\newcommand{\ptfcomma}{$p_T^f$}
\newcommand{\eqnptf}{p_T^f}
\newcommand{\pti}{$p_T^i$ }
\newcommand{\pticomma}{$p_T^i$}
\newcommand{\eqnpti}{p_T^i}
\newcommand{\lowpt}{low-\pt}
\newcommand{\lowptcomma}{low-\ptcomma}
\newcommand{\midpt}{mid-\pt}
\newcommand{\midptcomma}{mid-\ptcomma}
\newcommand{\intermediatept}{intermediate-\pt}
\newcommand{\intermediateptcomma}{intermediate-\ptcomma}
\newcommand{\highpt}{high-\pt}
\newcommand{\highptcomma}{high-\ptcomma}
\newcommand{\Aperp}{$A_\perp$ }
\newcommand{\Aperpcomma}{$A_\perp$}
\newcommand{\eqnAperp}{A_\perp}
\newcommand{\rperp}{$r_\perp$ }
\newcommand{\rperpcomma}{$r_\perp$}
\newcommand{\eqnrperp}{r_\perp}
\newcommand{\eqnrperpHS}{r_{\perp,HS}}
\newcommand{\eqnrperpWS}{r_{\perp,WS}}
\newcommand{\Rperp}{$R_\perp$ }
\newcommand{\Rperpcomma}{$R_\perp$}
\newcommand{\eqnRperp}{R_\perp}

\newcommand{\pizero}{$\pi^0$ }
\newcommand{\eqnpizero}{\pi^0}
\newcommand{\pizerocomma}{$\pi^0$}

\newcommand{\qhat}{$\hat{q}$ }
\newcommand{\qhatcomma}{$\hat{q}$}
\newcommand{\eqnqhat}{\hat{q}}

\newcommand{\gym}{$g_{SYM}$ }
\newcommand{\gymcomma}{$g_{SYM}$}
\newcommand{\eqngym}{g_{SYM}}
\newcommand{\gsym}{\gym}
\newcommand{\gsymcomma}{\gymcomma}
\newcommand{\eqngsym}{\eqngym}
\newcommand{\gs}{$g_{s}$ }
\newcommand{\gscomma}{$g_{s}$}
\newcommand{\eqngs}{g_{s}}
\newcommand{\asym}{$\alpha_{SYM}$ }
\newcommand{\asymcomma}{$\alpha_{SYM}$}
\newcommand{\eqnasym}{\alpha_{SYM}}
\newcommand{\alphasym}{\asym}
\newcommand{\alphasymcomma}{\asymcomma}
\newcommand{\eqnalphasym}{\eqnasym}

\newcommand{\stronglycoupled}{strongly-coupled }
\newcommand{\weaklycoupled}{weakly-coupled }
\newcommand{\stronglycoupledcomma}{strongly-coupled}
\newcommand{\weaklycoupledcomma}{weakly-coupled}
\newcommand{\ads}{AdS/CFT }
\newcommand{\asd}{\ads}
\newcommand{\adscomma}{AdS/CFT}
\newcommand{\asdcomma}{\adscomma}
\newcommand{\thooft}{'t Hooft }
\newcommand{\thooftcomma}{'t Hooft}

\newcommand{\infinity}{\infty}

\newcommand{\rightleftarrow}{\leftrightarrow}

\newcommand{\eq}[1]{Eq.~(\ref{#1})}
\newcommand{\eqn}[1]{Eq.~(\ref{#1})}
\newcommand{\fig}[1]{Fig.~\ref{#1}}
\newcommand{\figtwo}[2]{Figs.~\ref{#1}, \ref{#2}}
\newcommand{\tab}[1]{Table \ref{#1}}


\newcommand{\captionsize}{\small}